\documentclass[conference]{IEEEtran}

\newif\ifproposal
    \proposalfalse

\newif\ifthesis
    \thesisfalse

\newif\ifieeejournal
    \ieeejournalfalse
\newif\ifbiographies
    \biographiesfalse
\newif\ifreplytoreviewers
    \replytoreviewersfalse

\newif\ifieeeconference
    \ieeeconferencefalse

\newif\ifisf
    \isffalse

\newif\ifmicro
    \microfalse

    \ieeeconferencetrue

\usepackage{float}

\usepackage{basic_packages} 

\renewcommand{\eqref}[1]{(\ref{#1})}

\newcommand{\figref}[1]{\mbox{Fig.~\ref{#1}}}

\newcommand{\tblref}[1]{\mbox{Table~\ref{#1}}}

\definecolor{applegreen}{rgb}{0.55, 0.71, 0.0}
\usepackage{environ}
\newcounter{charcount}
\newcounter{charlim}
\NewEnviron{countenv}[2][]{%
  \setcounter{charcount}{0}%
  \expandafter\countem\BODY\relax\EOE
  \relax%
  \ifnum\value{charcount}<\value{charlim}\relax
      \begin{#2}\BODY\end{#2}
  \else
      \begin{#2}#1 \BODY  
         \color{red}\thecharcount{} letters in the #2 exceeds
         \thecharlim{} character limit. 
         Use the ``charlim'' parameter to change the character limit. \color{black}
       \end{#2} 
  \fi
}

\long\def\countem#1#2\EOE{%
  \stepcounter{charcount}%
  \ifx\relax#2
    \def\next{\relax}%
  \else
    \def\next{\countem#2\EOE}%
  \fi
  \expandafter\next%
}

\usepackage{scalerel}
\usetikzlibrary{svg.path}
\definecolor{orcidlogocol}{HTML}{A6CE39}
\tikzset{
  orcidlogo/.pic={
    \fill[orcidlogocol] svg{M256,128c0,70.7-57.3,128-128,128C57.3,256,0,198.7,0,128C0,57.3,57.3,0,128,0C198.7,0,256,57.3,256,128z};
    \fill[white] svg{M86.3,186.2H70.9V79.1h15.4v48.4V186.2z}
                 svg{M108.9,79.1h41.6c39.6,0,57,28.3,57,53.6c0,27.5-21.5,53.6-56.8,53.6h-41.8V79.1z M124.3,172.4h24.5c34.9,0,42.9-26.5,42.9-39.7c0-21.5-13.7-39.7-43.7-39.7h-23.7V172.4z}
                 svg{M88.7,56.8c0,5.5-4.5,10.1-10.1,10.1c-5.6,0-10.1-4.6-10.1-10.1c0-5.6,4.5-10.1,10.1-10.1C84.2,46.7,88.7,51.3,88.7,56.8z};
  }
}

\newcommand\orcidicon[1]{\href{https://orcid.org/#1}{\mbox{\scalerel*{
\begin{tikzpicture}[yscale=-1,transform shape]
\pic{orcidlogo};
\end{tikzpicture}
}{|}}}}
\newcommand{\X}{$\times$\xspace}    % times -> x
    
\newcommand{\mm}{\,\si{\mm}\xspace}

\newcommand{\m}{\,\si{\m}\xspace}

\newcommand{\nm}{\,\si{\nm}\xspace}

\newcommand{\um}{\,\si{\um}\xspace}

\newcommand{\s}{\,\si{\s}\xspace}
\newcommand{\ps}{\,\si{\ps}\xspace}
\newcommand{\ns}{\,\si{\ns}\xspace}
\newcommand{\us}{\,\si{\us}\xspace}
\newcommand{\ms}{\,\si{\ms}\xspace}

\newcommand{\bit}{\,\si{\bit}\xspace}

\newcommand{\Hz}{\,\si{\Hz}\xspace}
\newcommand{\kHz}{\,\si{\kHz}\xspace}
\newcommand{\MHz}{\,\si{\MHz}\xspace}
\newcommand{\GHz}{\,\si{\GHz}\xspace}

\newacronym{iot}{IoT}{internet-of-things}

\newacronym{snr}{SNR}{signal-to-noise ratio}
    
\newacronym{ppa}{PPA}{performance, power, and area}

\newacronym{cdf}{CDF}{cumulative distribution function}
\newacronym{pdf}{PDF}{probabililty distribution function}
\newacronym{ip}{IP}{intellectual property}
\newacronym{fir}{FIR}{finite impulse response}
\newacronym{dsp}{DSP}{digital signal processing}
\newacronym{lut}{LUT}{look-up table}
\newacronym{mac}{MAC}{multiply-and-accumulate}

\newacronym{dl}{DL}{deep learning}
    
\newacronym{ml}{ML}{machine learning}
    
\newacronym{ai}{AI}{artificial-intelligence}

\newacronym{cnn}{CNN}{convolutional neural network}
    
\newacronym{dnn}{DNN}{deep neural network}

\newacronym[longplural={graphic processing units}]{gpu}{GPU}{graphic processing unit}

\newacronym[longplural={central processing units}]{cpu}{CPU}{central processing unit}

\newacronym{tpu}{TPU}{tensor processing unit}
    
\newacronym{relu}{ReLu}{rectified linear unit}
    
\newacronym{llm}{LLM}{Large Language Model}

\newacronym{enics}{EnICS}{Emerging NanoScaled Integrated Circuits \& Systems}
    
    \newcommand{\enicsAffiliation}{EnICS Labs, Bar Ilan University, Ramat Gan 5290002, Israel\xspace} 

\newacronym{BIU}{BIU}{Bar-Ilan University\xspace}
     
\newacronym{UNICAL}{UNICAL}{University of Calabria\xspace}
    
\newacronym{DIMES}{DIMES}{Department of Computer Engineering, Modeling, Electronics and Systems\xspace}
    
\newacronym{USFQ}{USFQ}{Universidad San Francisco de Quito\xspace}
         
\newacronym{EPFL}{EPFL}{\'Ecole Polytechnique F\'ed\'erale de Lausanne\xspace}
     
\newcommand{\Mieee}{\IEEEmembership{, Member,~IEEE}} 
 
\newcommand{\SeMieee}{\IEEEmembership{, Senior Member,~IEEE}} 
 
\newacronym{isf}{ISF}{Israel Science Fund\xspace}
\newacronym{iia}{IIA}{Israel Innovation Authority\xspace}

\newacronym{itrs}{ITRS}{International Technology Roadmap for Semiconductors}
\newacronym{vlsi}{VLSI}{very large scale integration}
\newacronym{asic}{ASIC}{application specific integrated circuit}
\newacronym{pcb}{PCB}{printed circuit board}
\newacronym{cmos}{CMOS}{complementary-metal-oxide-semiconductor}

\newacronym[longplural={systems-on-chip}]{soc}{SoC}{system-on-chip}

\newacronym[longplural={integrated circuits}]{ic}{IC}{integrated circuit}

\newacronym{mc}{MC}{Monte Carlo}
     
\newacronym{mep}{MEP}{minimum energy point}

\newacronym[longplural={non-volatile memories}]{nvm}{NVM}{non-volatile memory}

\newacronym{vdd}{$V_{\text{DD}}$}{supply voltage}
    
\newacronym{gnd}{$GND$}{ground}
    
\newacronym{subvt}{sub-$V_{\text{T}}$}{sub-threshold}
    
\newacronym{nearvt}{near-$V_{\text{T}}$}{near threshold}
\newacronym{vt}{$V_{\text{T}}$}{threshold voltage}

\newacronym{vgs}{$V_{\text{GS}}$}{gate-to-source voltage} 
\newacronym{vds}{$V_{\text{DS}}$}{drain-to-source voltage} 
\newacronym{vbs}{$V_{\text{BS}}$}{body-to-source voltage} 
\newacronym{vgb}{$V_{\text{GB}}$}{gate-to-body voltage} 
\newacronym{dibl}{DIBL}{drain induced barrier lowering}
\newacronym{gidl}{GIDL}{gate induced drain leakage} 
\newacronym{ids}{$I_{\text{DS}}$}{drain-to-source current}
\newacronym{sce}{SCE}{short channel effect}
\newacronym{rsce}{RSCE}{reverse short channel effect}
\newacronym{tox}{$t_{\text{ox}}$}{gate oxide thickness}
\newacronym{L}{$L$}{channel length}
\newacronym{W}{$W$}{channel width}
\newacronym{rbb}{RBB}{reverse body biasing}
\newacronym{fbb}{FBB}{forward body biasing}
\newacronym{btbt}{BTBT}{band-to-band tunneling}
\newacronym{bjt}{BJT}{bipolar junction transistor}
\newacronym{hvt}{HVT}{high threshold voltage}
\newacronym{lvt}{LVT}{low threshold voltage}
\newacronym{nvt}{NVT}{nominal threshold voltage}
\newacronym{pmos}{PMOS}{p-type MOSFET}
\newacronym{nmos}{NMOS}{n-type MOSFET}
\newacronym{isub}{$I_\text{sub}$}{sub-threshold leakage}
\newacronym{igate}{$I_\text{gate}$}{gate leakage}
\newacronym{ibulk}{$I_\text{bulk}$}{bulk leakage}
\newacronym{vbb}{$V_{\text{BB}}$}{body voltage} 
    
\newacronym{ptm}{PTM}{predictive technology model}

\newacronym{pdk}{PDK}{process design kit}

\newacronym{sc}{SC}{standard cell}
\newacronym{vtc}{VTC}{voltage transfer characteristic}
\newacronym{dff}{DFF}{Data Flip-Flop}
\newacronym{dcvsl}{DCVSL}{differential cascade voltage switch logic}

\newacronym{dif}{DIF}{digital implementation flow}
\newacronym{hdl}{HDL}{hardware description language}
\newacronym{rtl}{RTL}{register transfer level}
\newacronym{eda}{EDA}{electronic design automation}
\newacronym{cad}{CAD}{computer-aided design}
\newacronym{pr}{P\&R}{place and route}
\newacronym{cts}{CTS}{clock-tree synthesis}
\newacronym{sta}{STA}{static timing analysis}
\newacronym{edi}{EDI}{Cadence Encounter Design Implementation}
\newacronym{s_dc}{DC}{Synopsys Design Compiler}
\newacronym{sdc}{SDC}{Synopsys Design Constraints}
\newacronym{vcd}{VCD}{value change dump}
\newacronym{pvt}{PVT}{Process-Voltage-Temperature}
\newacronym{scm}{SCM}{standard cell memory}
    
\newacronym{ips}{IPS}{instructions per second}

\newacronym{eflash}{eFlash}{embedded Flash}

\newacronym[longplural={Storage Class Memories}]{scmems}{SCM}{Storage Class Memory}

\newacronym{ddr}{DDR}{dual-data rate}

\newacronym{sata}{SATA}{Serial Advanced Technology Attachment}
    
\newacronym{nvme}{NVMe}{Non-Volatile Memory Express}
    
\newacronym{pcie}{PCIe}{Peripheral Component Interconnect Express}
     
\newacronym[longplural={hard-Disk drives}]{hdd}{HDD}{hard-Disk drive}

\newacronym[longplural={solid-State drives}]{ssd}{SSD}{solid-State drive}

\newacronym[longplural={high-bandwidth memories}]{hbm}{HBM}{high-bandwidth memory}

\newacronym[longplural={dual-inline memory modules}]{dimm}{DIMM}{dual-inline memory module}

\newacronym[longplural={dynamic random-access memories}]{dram}{DRAM}{dynamic random-access memory}

\newacronym{fifo}{FIFO}{first-in first-out}
\newacronym{lifo}{LIFO}{last-in first-out}
\newacronym[longplural={content addressable memories}]{cam}{CAM}{content addressable memory}
    
\newacronym{L1}{L1}{level-1}
\newacronym{L2}{L2}{level-2}
\newacronym{L3}{L3}{level-3}
\newacronym{L4}{L4}{level-4}
\newacronym{simd}{SIMD}{single-instruction multiple-data}

\newacronym{bc}{BC}{bitcell}

\newacronym{bl}{BL}{bitline}

\newacronym{sln}{SL}{sourceline}

\newacronym{wl}{WL}{wordline}

\newacronym[longplural={Gain-Cell embedded DRAMs}]{gcedram}{GC-eDRAM}{Gain-Cell embedded DRAM}

\newacronym{sixt}{6T}{6-transistor}
    
\newacronym[longplural={static random-access memories}]{sram}{SRAM}{static random-access memory}

\newacronym[longplural={six-transistor static random access memories}]{sixtsram}{6T-SRAM}{six-transistor static random access memory}

\newacronym[longplural={embedded DRAMs}]{edram}{eDRAM}{embedded DRAM}

\newacronym[longplural={multi-level cells}]{mlc}{MLC}{multi-level cell}

\newacronym{mw}{MW}{write transistor}
\newacronym{mr}{MR}{read transistor}
\newacronym{sn}{SN}{storage node}

\newacronym{wwl}{WWL}{write word line}

\newacronym{rwl}{RWL}{read word line}

\newacronym{sa}{SA}{sense amplifier}
\newacronym{drv}{DRV}{data retention voltage}
\newacronym{nwl}{NWL}{negative word line}
\newacronym{bist}{BIST}{built-in self-test}
\newacronym{bisr}{BISR}{built-in self-repair}
\newacronym{ecc}{ECC}{error correction code}
\newacronym{snm}{SNM}{static noise margin}
\newacronym{rsnm}{RSNM}{read static noise margin}
\newacronym{wsnm}{WSNM}{write static noise margin}
\newacronym{dnm}{DNM}{dynamic noise margin}
\newacronym{drt}{DRT}{data retention time}

\newacronym{lrs}{LRS}{low resistance state}

\newacronym{hrs}{HRS}{high resistance state}

\newacronym[longplural={phase-change memories}]{pcm}{PCM}{phase-change memory}

\newacronym[longplural={resistive RAMs}]{rram}{RRAM}{resistive RAM}

\newacronym{stt}{STT}{spin-transfer torque}

\newacronym[longplural={spin-transfer torque magnetic random-access memories}]{sttmram}{STT-MRAM}{spin-transfer torque magnetic random-access memory}

\newacronym[longplural={magnetic random-access memories}]{mram}{MRAM}{magnetic random-access memory}

\newacronym{mtj}{MTJ}{magnetic tunnel junction}

\newacronym{smtj}{SMTJ}{single-barrier MTJ}

\newacronym{dmtj}{DMTJ}{double-barrier MTJ}

\newacronym{mim}{MIM}{metal-insulator-metal}

\newacronym{euv}{EUV}{extreme ultra-violet}
     
\newacronym{soi}{SOI}{silicon-on-insulator}
\newacronym{fdsoi}{FD-SOI}{fully-depleted silicon-on-insulator}
    
\newacronym{rdf}{RDF}{random dopant fluctuations}
\newacronym{ocv}{OCV}{on-chip variation}

\newacronym{lpa}{LPA}{Leakage Power Analysis}
\newacronym{dpa}{DPA}{Differential Power Analysis}
\newacronym{puf}{PUF}{Physical Unclonable Function}

\newacronym{ser}{SER}{soft errors}
     
\newacronym{seu}{SEU}{single-event upset}
     
\newacronym{qcrit}{$Q_\text{crit}$}{critical charge}
\newacronym{tmr}{TMR}{triple modular redundancy}
\newacronym{dmr}{DMR}{dual modular redundancy}
\newacronym{edac}{EDAC}{error detection and correction}
\newacronym{secded}{SECDED}{single error correction~-- double error detection}
\newacronym{dected}{DECTED}{double error correction~-- triple error detection}

\newacronym{smu}{SMU}{source/measure unit}
\newacronym{dmm}{DMM}{digital multimeter}
\IEEEoverridecommandlockouts

\title{Retrospective: A CORDIC Based Configurable Activation Function for NN Applications}

 \author{
    \IEEEauthorblockN{
        Omkar Kokane\IEEEauthorrefmark{1}\orcidicon{0009-0000-6288-7231}, 
        Gopal Raut\IEEEauthorrefmark{1} \orcidicon{0000-0002-1046-9457}\Mieee,
        Salim Ullah\IEEEauthorrefmark{2}, \orcidicon{0000-0002-9774-9522}, Mukul Lokhande\IEEEauthorrefmark{1}\IEEEauthorrefmark{3}\orcidicon{0009-0001-8903-5159},\\
        Adam Teman\IEEEauthorrefmark{3}\orcidicon{0000-0002-8233-4711}\SeMieee,
        Akash Kumar\IEEEauthorrefmark{2}\orcidicon{0000-0001-7125-1737}\SeMieee,\\ and Santosh Kumar Vishvakarma\IEEEauthorrefmark{1}\orcidicon{0000-0003-4223-0077}\SeMieee
        }
        
    \IEEEauthorblockA{\IEEEauthorrefmark{1}NSDCS Research Group, Indian Institute of Technology Indore 453552, India}
    \IEEEauthorblockA{\IEEEauthorrefmark{2}Ruhr University Bochum D-44801, Germany}
    \IEEEauthorblockA{\IEEEauthorrefmark{3}\enicsAffiliation}
    Email: skvishvakarma@iiti.ac.in \textbf{(Corresponding Author)}\\
    \thanks{}   
}

\begin{document}
\maketitle

%----------ABSTRACT-------------------%
%-------------------------------------%
%\begin{abstract}

\begin{countenv}{abstract} 
% CORDIC-based configurable AF proposed to accelerate the ASIC hardware design and resource-constrained solution with functional reconfigurability. Since published in ISVLSI'20~\cite{Gopal-ISVLSI}, this new approach in neural network accelerators become pretty popular and spawned in many activation function designs, both in academia and commercial AI chips. In this retrospective, we review the background behind this project, summarize several works from the past few years and present DA-VINCI AF for emerging AI workloads. The Next-Gen dynamically configurable and precision-scalable activation function core extends flexibility for more diverse activation functions in AI workloads such as Swish, SoftMax, SELU, and GELU with the Shift-and-Add CORDIC methodology. The prior block has been optimized for MAC, Sigmoid, and Tanh functionality and incorporated into ReLU AF. These improvements make NEURIC a fundamental block in resource-efficient Vector Engine for AI accelerators focused on DNNs, RNNs/LSTMs, and Transformers, achieving a quality of results (QoR) of 98.5\%. 

A CORDIC-based configuration for the design of Activation Functions (AF) was previously suggested to accelerate ASIC hardware design for resource-constrained systems by providing functional reconfigurability. Since its introduction, this new approach for neural network acceleration has gained widespread popularity, influencing numerous designs for activation functions in both academic and commercial AI processors. In this retrospective analysis, we explore the foundational aspects of this initiative, summarize key developments over recent years, and introduce the DA-VINCI AF tailored for the evolving needs of AI applications. This new generation of dynamically configurable and precision-adjustable activation function cores promise greater adaptability for a range of activation functions in AI workloads, including Swish, SoftMax, SeLU, and GeLU, utilizing the Shift-and-Add CORDIC technique. The previously presented design has been optimized for MAC, Sigmoid, and Tanh functionalities and incorporated into ReLU AFs, culminating in an accumulative NEURIC compute unit. These enhancements position NEURIC as a fundamental component in the resource-efficient vector engine for the realization of  AI accelerators that focus on DNNs, RNNs/LSTMs, and Transformers, achieving a quality of results (QoR) of 98.5\%.

\end{countenv}
%\end{abstract}


%----------INTRODUCTION---------------%
%-------------------------------------%
\section{Configurable CORDIC Approach \& Growth}
\label{sec_introduction}

The past few decades have seen a rise in the use of AI applications across different domains, such as DNN inference and training, Vision Transformers (ViTs), Generative AI, RNNs, LSTMs, etc. Thus, the need for accelerated computing has increased significantly in executing computationally complex models. The major focus has always been on optimizing the computing units (CUs) for GEMM, MVM or MAC computations. However, the other workloads, including the non-linear activation functions (AFs), have become significant bottlenecks in the execution~\cite{CORDICAF-RNN, Flex-PE}. 
To address different AF requirements, our work in ISVLSI'20~\cite{Gopal-ISVLSI} proposed an efficient yet configurable approach for AF design based on CORDIC. 
This project started with the design of ASIC accelerators that required configurable non-linear transfer functions in ANNs, which were earlier limited to FPGAs. 
Previous work~\cite{GR-ACM_TRETS23, AF-reusedNeuro22, Gopal-ISVLSI, Intel-taylor_Patent23, NVIDIA-SoftMax_Patent22, Intel-ApproxLUT_Patent24_1, Intel-ApproxLUT_Patent23_2} primarily focused on LUT-based approaches that store values or parameters, LUT-based piecewise linear (PWL) approximation, Stochastic computation (SC) techniques, Taylor series approximation~\cite{ReAFM-NN}, etc. 
Our goal was to scale the precision capability while maintaining reconfigurability. This led to the development of ``RECON''~\cite{RECON}, a technique to reduce the excess burden of area requirement of non-linear activation functions with MAC hardware and scalable design at 8/16-bits. 
The resulting unified data path was more straightforward and more area-efficient for DNN accelerators. 
This proved that special-purpose hardware could be cost-effective. 
This not only saves static power consumption during AF computation but also allows for a reduction in the effective dark-silicon area. 
This demonstrates the opportunity for precision-scalable and reconfigurable SIMD accelerators. 

\begin{table*}[!t]
    \caption{Comparison of state-of-the-art Activation Function design methodologies.}
    \label{tab:gen-comp}
    \renewcommand{\arraystretch}{1.25}
    \resizebox{\textwidth}{!}{%
    \begin{tabular}{c|c|c|c|c|c|c|c|c|c}
    \hline
    \textbf{} & \textbf{Shen et al.}~\cite{CORDICAF-RNN}  &  \textbf{Yang et al.}~\cite{Designspaceexploration-AF} & \textbf{AFB}~\cite{CORDICAF-LSTM} & \textbf{Mahati et al.}~\cite{MRao-ISQED24} & \textbf{SoftAct}~\cite{SoftAct-Trans}, \textbf{AxSF}~\cite{SoftMax-taylor-DNN} & \textbf{Hong et al.}~\cite{RECONFIG-MP-QuantAware-NAF} & \textbf{RECON}~\cite{RECON} & \textbf{ReAFM}~\cite{ReAFM-NN}  & \textbf{DA-VINCI} \\ \hline
    \textbf{Precision} & - & 8-bit & FXP16 & FP16/32, BF16 & 16-bit  & Custom FP & FXP8/16/32 & 12-bit  & FXP 8/16 \\ \hline
    \textbf{Design approach} & CORDIC & PWL-LUT & CORDIC & CORDIC & Approximation  & LUT & CORDIC & Approximation  & CORDIC \\ \hline
    \textbf{Non-linear AF} & Sigmoid, Tanh & Tanh, SELU & Sigmoid, Tanh & \begin{tabular}[c]{@{}c@{}}SoftMax\\Sigmoid, Tanh\end{tabular} & SoftMax  & \begin{tabular}[c]{@{}c@{}}Swish\\Sigmoid, Tanh\end{tabular}   
    & Sigmoid, Tanh & 
    \begin{tabular}[c]{@{}c@{}}Swish\\GELU, ReLU \end{tabular}  & \begin{tabular}[c]{@{}c@{}}Swish, SoftMax, SeLU\\GELU, Sigmoid, Tanh, ReLU\end{tabular} \\ \hline
    \textbf{Design flexibility} & Yes & No & Yes & No & No  & Yes & Yes & Yes  & Yes \\ \hline
    \textbf{AI workloads} & RNN & DNN & LSTM & - & DNN/Transformers  & - & DNN & DNN & DNN, RNN, LSTM, ViTs \\ \hline
    \end{tabular}}
\end{table*}

That said, plenty of scope remains to enable more comprehensive functionality with the same hardware for accelerating diverse AI workloads. 
For example, the Google TPUv4 allocated 20\% of the area for less than 2\% of the non-linear functions for popular neural networks~\cite{Tiny-SoftMax-TCASII'23}, which emphasizes the need for better hardware implementation of AFs.

Since the original design was proposed, several AF accelerators based on CORDIC have been presented. 
\cite{CORDICAF-RNN} proposed CORDIC-based Sigmoid/Tanh for RNN design targeting FPGA-based acceleration.
\cite{MRao-ISQED24} utilizes CORDIC-based Exponential and Division units for Tanh/Sigmoid/SoftMax activation functions in FP32, FP16/BF16/TF32, and Posit16 precision.
\cite{GR-ISQED24} optimized SIMD dynamic fixed-point CORDIC design for 8/16 bit for Tanh/Sigmoid.
\cite{TCASI23-Softmax} proposed CORDIC-based SoftMax for faster DNN training optimization.
\cite{GR-ACM_TRETS23} generalizes CORDIC architecture for reusable MAC, Sigmoid/Tanh.
\cite{HOAA} enhances the operating frequency of CORDIC-Neuron/PE.
\cite{SoftMax-Sumi} proposed precision-scalable FxP8/16/32 SoftMax for DNN applications.
\cite{Flex-PE} proposed SIMD and reconfigurable FxP4/8/16/32 CORDIC-based processing elements (PEs), capable of MAC~\cite{Quant-PE}, Sigmoid, Tanh and SoftMax computations. 
This includes an iterative version for edge devices and a pipelined version for HPC-cloud devices. 
The detailed comparison of prior works~\cite{DL-AF-Survey, VisionTrans-AF, GenAI-LLMs-AF, Designspaceexploration-AF, CORDICAF-RNN, CORDICAF-LSTM, TCASI23-Softmax, SoftAct-Trans} with their hardware approaches for acceleration of various workloads is provided in \tblref{tab:gen-comp}. 

CORDIC also had substantial impact on commercial AI hardware design. 
AMD Xilinx FPGAs (UltraScale/Versal Series 2022/23) support optimized CORDIC hardware IP blocks\cite{Xil-IP, GR-ACM_TRETS23}. 
Intel DSP builder 2022 adopted the CORDIC approach in cases of the absence of faster LUT blocks in the HLS approach\cite{Intel-DSP}. 
STMicro SR5E1x family integrates a 32-bit Arm-based Cortex-M7 with a specialized CORDIC accelerator, enabling hardware-level support for trigonometric math required in motor control, metering, and signal processing, thus significantly enhanced software performance\cite{STM-CORDIC}. NI and Mathworks framework support CORDIC-based math libraries for trigonometric and hyperbolic functions with parameterizable data width for execution, using either a serial architecture for minimal area implementations or a parallel architecture for speed optimization.

\section{Lessons Learned \& Need of the Hour}

Despite the fact that CORDIC started with AF acceleration, this technique is not as easily applied to different workloads. 
SoftMax requires additional FIFO~\cite{SoftMax-Sumi} overhead. 
Several other designs enhanced the performance, including HOAA-enabled PE~\cite{HOAA} and radix-4 CORDIC~\cite{radix4CORDIC} trading off hardware efficiency and error rate.
Another improvement is precision-configurable Flex-PE~\cite{Flex-PE} to enhance the quantized throughput in addition to reconfigurability.  
The CORDIC special-purpose hardware is orders of magnitude more efficient than a software implementation of pseudo-CORDIC with ALU; the overhead of multiple RISC-V instructions suffer from significant delay. 
One ALU can calculate a CORDIC MAC within 135\X{2} cycles to implement Shift and Add/Sub instructions. 
Other designs use a PWL-based approach to reduce control overhead. 

Configurable AF only accelerates fully connected layers.
CORDIC-Neuron can also accelerate convolution layers. 
Commercially, Xilinx CORDIC IP tried to apply this approach on everything. 
While it was shown to be perfect for vision models, it is not appropriate for LLMs, since the data flow of recent LLMs becomes more complex. 
The design of reconfigurable MAC and SoftMax hardware to accelerate transformers is complex. 

This paper addresses the aforementioned issues with dynamically configurable and runtime precision-scalable activation functions in versatile neurons via CORDIC implementation (DA-VINCI). 
DA-VINCI could be considered a fundamental block in AI hardware design. 
The specific contributions of this work are summarized as follows: 

\begin{itemize}
    \item \textbf{Novel DA-VINCI programmable AF core:}
    The proposed DA-VINCI core supports activation functions with almost negligible overhead. 
    The activation functions support runtime reconfigurability between Swish, SoftMax, SELU, GELU, Sigmoid, Tanh, and ReLU.

    \item \textbf{NEURIC – Neuron Engine based on Reconfigurable Iterative CORDIC core:}
    The DA-VINCI-based NEURIC core has been proposed with a reconfigurable 8/16-bit precision MAC + AF unit. 
    This serves as a fundamental block in the realization of lightweight reconfigurable layer-multiplexed AI cores.  

    \item \textbf{Empirical approach for enhanced AI throughput:}
    The proposed NEURIC PE is evaluated for reconfigurable 8/16-bit precision and analyzed for a layer-multiplexed vector engine. 
    It has also been benchmarked for several industry-standard benchmarks and proven to be within 98.5\% Quality of results (QoR). 
    The CORDIC-based shift-add approach provides versatility, which has been analyzed in-depth with FPGA and ASIC validation. 
\end{itemize}

\begin{figure*}[!t]
    \centering
    \includegraphics[width=1.8\columnwidth]{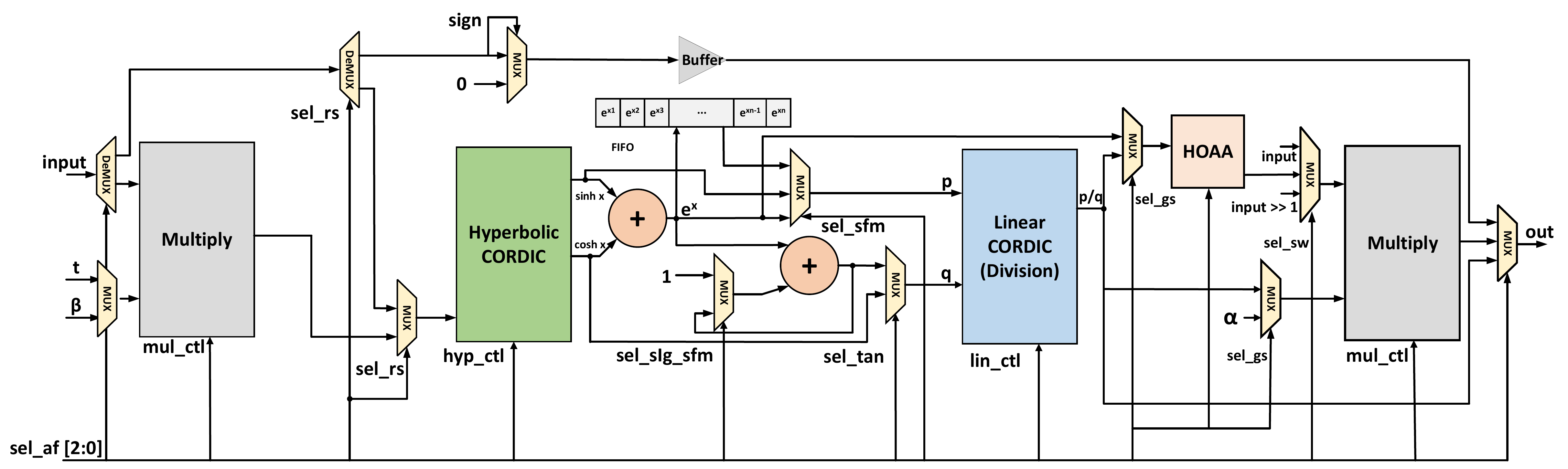}
    \vspace{-3mm}
    \caption{The detailed micro-architecture of DA-VINCI core with integrated data flow and control signals}
    \vspace{-3mm}
    \label{smart-af}
\end{figure*}

%%%%%%%%%%%%%%%%%%%%%%%%%%%%%%%%%%%%%%%%%%%%%%%%%%%%%%%%%%%
\section{New approach and AI Evaluation}
\label{sec_proposed_solution}

Neural network architectures have gone through rapid changes over the recent past. 
From DNN to RNN/LSTM to Transformers, functional-aware hardware design techniques are needed for a compact and fast acceleration approach. 
The first principle of efficient AI computing is to be lazy: avoid data movement, quickly reconfigure the workload, or combine the functionality. 

The fundamental processing element in AI hardware comprises MAC and AF units~\cite{HOAA}. 
Several methodologies focus on reconfigurability for enhanced compute density~\cite{RASHT-Reconfigurable-NN, RECON, Unified-CORDIC, GR-ACM_TRETS23}. 
Many approaches have been proposed in the past few years, and their impacts on application performance are summarized below. 

In~\cite{Designspaceexploration-AF, Intel-ApproxLUT_Patent23_2, Intel-ApproxLUT_Patent24_1}, the function values are stored in LUTs; however, this approach is not precision-scalable, as significantly high storage and latency come into the picture at higher precision. 
The approach presented in~\cite{RECONFIG-MP-QuantAware-NAF, ReAFM-NN} stores the slope and intercept of the function to compute, which introduces a minor enhancement. 
Taylor series approximation~\cite{SoftMax-taylor-DNN, Intel-taylor_Patent23} requires several multiplication and clock cycles. 
Another approximation with a power-of-2 requires less hardware, but the latency is still high. 
In addition, approximation methodologies impact performance due to interpolation and granularity issues. 
The application performance is severely degraded due to the step artifacts and quantization errors introduced by segment selection and convergence terms in the piecewise linear (PWL) approximation, stochastic computation (SC), and Taylor series approximation approaches.
These approaches are not very hardware-friendly, and thus, shift-add-based CORDIC~\cite{Unified-CORDIC, CORDICAF-RNN, CORDICAF-LSTM, MRao-ISQED24, RECON} has emerged as a recent trend for resource-efficient activation function design. 
Thus, we have chosen a performance-enhanced CORDIC approach for this work~\cite{HOAA}, which helps our design to be significantly better than SoTA CORDIC solutions. 

\begin{table}
    \caption{Comparison of FPGA Resources for SoTA AF Designs}
    \label{tab:FPGA-comp}
    \renewcommand{\arraystretch}{1.25}
    \resizebox{\linewidth}{!}{%
    \begin{tabular}{c|c|c|c|c|c}
    \hline
    \textbf{Parameter} & \textbf{ReAFM~\cite{ReAFM-NN}} & \textbf{Shen et al.~\cite{CORDICAF-RNN}} & \textbf{AFB~\cite{CORDICAF-LSTM}} & \textbf{AxSF~\cite{SoftMax-taylor-DNN}} & \textbf{DA-VINCI} \\ \hline
    
\textbf{\begin{tabular}[c]{@{}c@{}}AFs \\ Supported\end{tabular}} & \begin{tabular}[c]{@{}c@{}}Swish, Tanh,\\ Sigmoid\end{tabular} & \begin{tabular}[c]{@{}c@{}}Sigmoid,\\ Tanh\end{tabular} & \begin{tabular}[c]{@{}c@{}}Sigmoid,\\ Tanh\end{tabular} & SoftMax & \begin{tabular}[c]{@{}c@{}}SoftMax, Sigmoid, Tanh,\\ ReLU, GeLU, SeLU, Swish\end{tabular} \\ \hline
    \textbf{Precision} & 12-bit & - & 16-bit & 16-bit & 8/16-bit \\ \hline
    \textbf{Mean error (\%)} & 2.2 & 3.4 & 2.8 & - & 2.78 \\ \hline
    \textbf{Evaluation Board} & Virtex-7 & Zynq-7 & PYNQ-Z1 & Zynq-7 & Zybo \\ \hline
    \textbf{LUTs} & 367 & 2395 & 36286 & 1215 & 537 \\ \hline
    \textbf{FFs} & 298 & 1503 & 24042 & 1012 & 468 \\ \hline
    \textbf{Delay ($\mu$s)} & 0.35 & 0.18 & 21 & 3.32 & 2.56 \\ \hline
    \textbf{Power (mW)} & - & 0.681 & 125 & 165 & 30.38 \\ \hline
    \end{tabular}}
    \vspace{-3mm}
\end{table}

\begin{table}
    \caption{Comparison of ASIC Parameters for SoTA AF Designs}
    \label{tab:ASIC-comp}
    \renewcommand{\arraystretch}{1.25}
    \resizebox{\linewidth}{!}{%
\begin{tabular}{c|c|c|c|c|cc}
\hline
\textbf{Parameter} & \textbf{Zhang et al.~\cite{TCASI23-Softmax}} & \textbf{AxSF~\cite{SoftMax-taylor-DNN}} & \textbf{AFB~\cite{CORDICAF-LSTM}} & \textbf{RECON~\cite{RECON}} & \multicolumn{2}{c}{\textbf{DA-VINCI}} \\ \hline
\textbf{\begin{tabular}[c]{@{}c@{}}AFs \\ Supported\end{tabular}} & SoftMax & SoftMax & \begin{tabular}[c]{@{}c@{}}Sigmoid,\\ Tanh\end{tabular} & \begin{tabular}[c]{@{}c@{}}Sigmoid,\\ Tanh\end{tabular} & \multicolumn{2}{c}{\begin{tabular}[c]{@{}c@{}}SoftMax, Tanh, Sigmoid\\ ReLU, SeLU, GeLU, Swish\end{tabular}} \\ \hline
\textbf{Precision} & 32-bit & 16-bit & 16-bit & 16-bit & \multicolumn{2}{c}{8/16-bit} \\ \hline
\textbf{Tech. Node (nm)} & 28 & 28 & 45 & 45 & \multicolumn{1}{c|}{45} & 28 \\ \hline
\textbf{Area ($\mu$m\textsuperscript{2})} & 98787 & 3819 & 870523 & 24608 & \multicolumn{1}{c|}{6107} & 2138 \\ \hline
\textbf{Delay (ns)} & 26 & 1.6 & - & 4.76 & \multicolumn{1}{c|}{3.63} & 2.6 \\ \hline
\textbf{Power (mW)} & 24.72 & 1.58 & 151 & 1033 & \multicolumn{1}{c|}{72.3} & 59.8 \\ \hline
\end{tabular}}
\vspace{-3mm}
\end{table}

\subsection{DA-VINCI programmable AF Core}

The unified CORDIC algorithm~\cite{RECON, Unified-CORDIC} is preferred for reconfigurable hardware for circular, linear, and hyperbolic operations combined with rotational and vector modes. 
The fundamental CORDIC hardware uses simple design elements such as Add/Sub, MUX, logarithmic barrel shifter (LBS), and memory blocks. 
The HOAA-enabled CORDIC architecture~\cite{HOAA} provides 21\% area savings and up to 33\% lower power consumption, effectively compensating for the area overhead introduced with DA-VINCI core's reconfigurability. 
Thus, the proposed DA-VINCI core almost matches the resource utilization for individual AF implementations.  
The detailed methodology for CORDIC-based activation function hardware has been explored for Swish~\cite{ReAFM-NN}, SoftMax~\cite{SoftAct-Trans}, SELU~\cite{Designspaceexploration-AF}, GELU~\cite{ReAFM-NN}, Sigmoid~\cite{CORDICAF-LSTM}, Tanh~\cite{CORDICAF-RNN}, and ReLU.

The resource-constrained CORDIC implementation for arithmetic, trigonometric and complex mathematical functions can be easily understood with the help of mathematical equations of pseudo-rotation:
\begin{equation}
    \begin{aligned}
    X_{i+1} &= X_i - m \cdot d_i \cdot Y_i \cdot 2^{-i} \\
    Y_{i+1} &= Y_i + d_i \cdot X_i \cdot 2^{-i} \\
    Z_{i+1} &= Z_i - d_i \cdot E_i
    \end{aligned}
    \label{CORDIC-algo}
\end{equation}

The variables converge to $X_i$, $Y_i$, and $Z_i$, after the $i^{\text{th}}$ iteration.$ ( E_i )$ is $( 2^{-i} )$, $( \tan^{-1}(2^{-i}) )$, and $( \tanh^{-1}(2^{-i}) )$, and mode $( m \in \{0, 1, -1\} )$ depend on Linear, Circular, and Hyperbolic coordinates, according to the iteration cycle. The operation mode defines the scaling factor and rotation direction based on combinatorial logic. 
The precision  is limited by the CORDIC convergence range; for example, Hyperbolic Rotational (HR) mode \([-1.1182, 1.1182]\), Linear Vector (LV) mode \([-1, 1]\), and LR mode \([-7.968, 7.968]\). 
However, our approach has focused on normalized computations in the range of  \([-1, 1]\) with MaxNorm 5.5, thus ensuring output convergence. 

\begin{figure}[!b]
    \centering
    \includegraphics[width=\columnwidth]{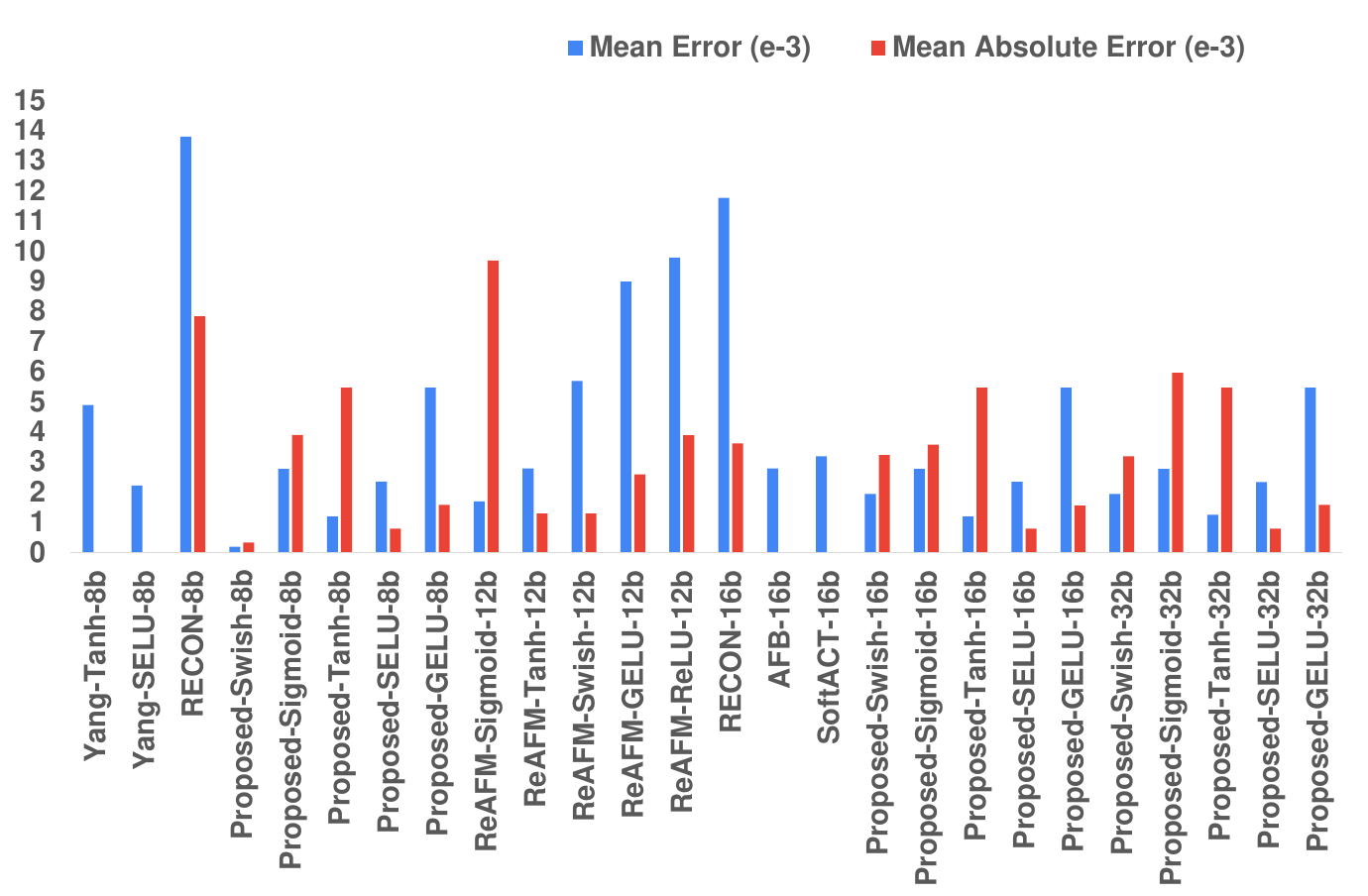}
    \vspace{-3mm}
    \caption{Comparison of error estimation (compared to baseline FP32) with state-of-the-art works~\cite{Designspaceexploration-AF, RECON, ReAFM-NN, CORDICAF-LSTM, SoftAct-Trans}.}
    \vspace{-2mm}
    \label{fig:smart-af-error}
\end{figure}

The core component of the proposed DA-VINCI programmable core includes
linear CORDIC and hyperbolic CORDIC units, as shown in \figref{smart-af}. 
Based on the \texttt{sel\_af} signal, the programmable core can activate a particular activation function with appropriate datapath selection and operation sequencing. 
HR mode is used to compute the $\sinh$ and $\cosh$ functions for Swish, SoftMax, SELU, GELU, Sigmoid and Tanh with 86\% hardware utility factor. 
LV mode is used for division operation in Swish, SoftMax, GELU, Sigmoid, and Tanh with a 72\% hardware utility factor. 
The additional area overhead includes a mux for operational switching between Sigmoid/Tanh, a buffer for ReLU, a
FIFO to store values in SoftMax computations and two multipliers in GELU implementation, while enabling additional SELU and Swish within this extra overhead, which justifies the reconfigurability of the proposed design within minimal dark-silicon overhead. 
%%%%%%%%%%%%%%%%%%%%%%%%%%%%%%%%%%%%%%%%%%%%%%%%%%%%%%%%%%%%%%%%%%

\begin{figure}[!t]
    \centering
    \includegraphics[scale=0.18]{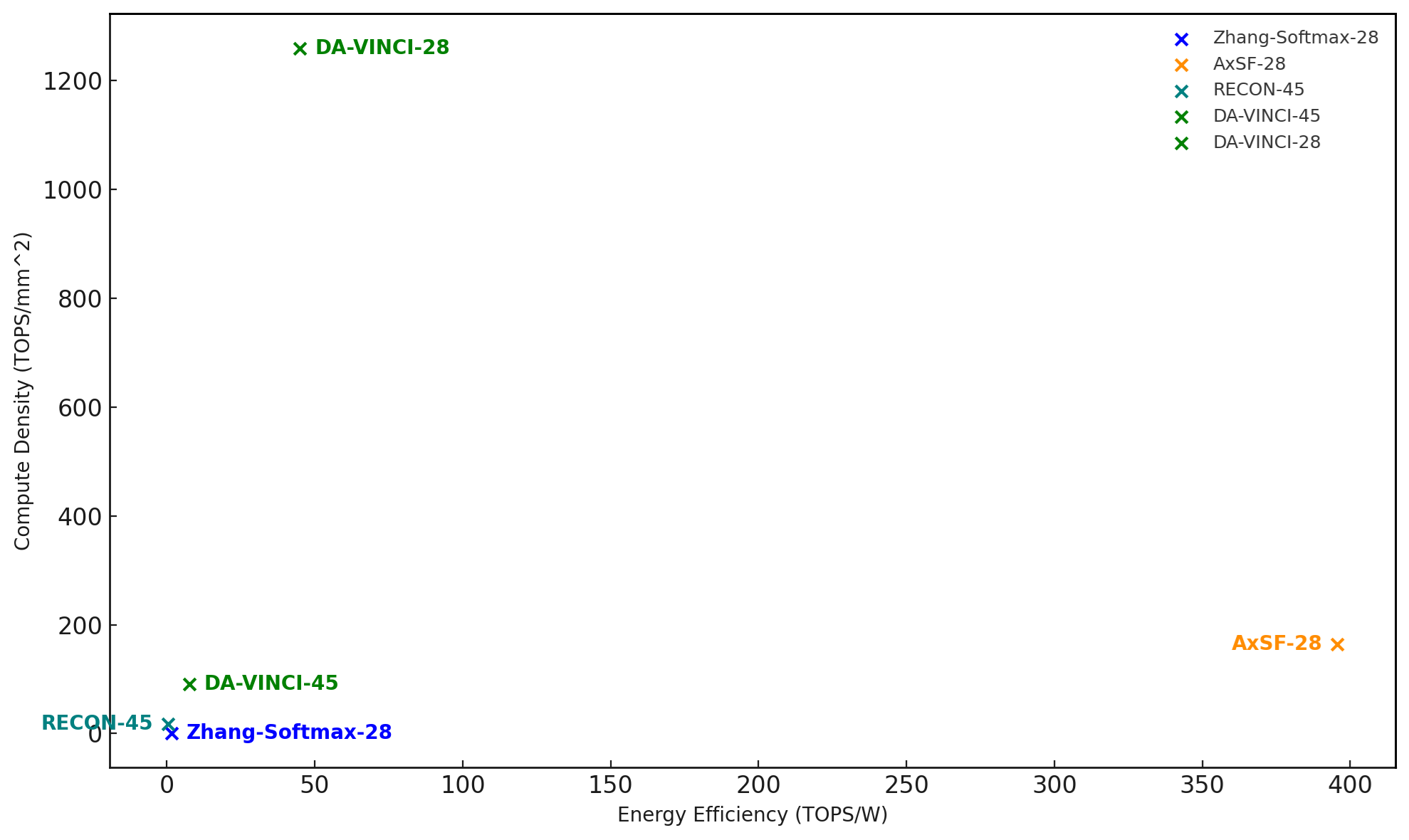}
    \vspace{-2mm}
    \caption{Performance analysis (ASIC: Energy Efficiency vs Compute Density) with state-of-the-art works~\cite{TCASI23-Softmax, RECON, SoftMax-taylor-DNN}.}
    \label{fig:perf-comp}
    \vspace{-5mm}
\end{figure}

We developed the experimental setup to empirically evaluate the proposed DA-VINCI programmable core, focusing on software-hardware architectural emulation. 
The proposed CORDIC-based model was designed with Python 3.0, FxP-math and QKeras 2.4 library for the Jupyter Notebook platform at 8/16/32-bit precision to validate arithmetic correctness. 
The outputs were stored in a text file to validate the UART-enabled NEURIC DUT. 
The Python error-estimation model was Monte Carlo simulated for uniformly distributed random-input patterns, and error estimates are reported in \figref{fig:smart-af-error} with respect to the true Python Numpy outputs. The optimal iterative stages for enhanced hardware performance are derived from the Pareto analysis. The application performance (accuracy) was evaluated for the diverse AI models, incorporating the proposed CORDIC-based AF unit across various datasets and are reported in \figref{perf-eval}. 
The software evaluation reveals that the proposed design achieves satisfactory performance within 98.5\% QoR for VGG-16 and ResNet-50 using CIFAR-100, MobileViT using ImageNet, etc. 

\subsection{NEURIC Neuron Evaluation}

The proposed NEURIC design is implemented with System-Verilog language at SIMD 8/16 FxP precision, with the FSM, based on an iterative or pipelined execution strategy. The design is reconfigurable MAC and DA-VINCI AF design with underlined CORDIC hardware.
The output simulation is functionally validated with Synopsys VCS and was found to be equivalent to Python emulation results.
Furthermore, FPGA synthesis and implementation was carried out with the help of the AMD Vivado Design Suite, and the utilization of post-implementation resources is reported in the Xilinx Zybo FPGA Evaluation Kit (XC7z010) in \tblref{tab:FPGA-comp}.
Different processing elements are compared with proposed CORDIC-approach in \tblref{table:comp-PE}: Ac2Ap2-PE~\cite{Acc-App-PE}, QuantPE~\cite{Quant-PE}, Xilinx-IP (Xil-PE)~\cite{GR-ACM_TRETS23} supports limited Sigmoid/Tanh,  while FP-MPE, FP-RPE~\cite{Reconfigurable-PE-TCASII24} supports ReLU only.
Compared with state-of-the-art designs, the proposed design reduces LUTs up to 4.5$\times$ and 2.3$\times$, and FFs up to 3.2$\times$ and 2.2$\times$ compared to SoTA works~\cite{CORDICAF-RNN} and~\cite{SoftMax-taylor-DNN}, respectively. 
Furthermore, the evaluation also shows an improvement in critical delay up by 30\% compared to~\cite{SoftMax-taylor-DNN} and a reduction in power consumption up to 4.2$\times$ and 5.4$\times$ compared to SoTA works~\cite{CORDICAF-LSTM} and~\cite{ SoftMax-taylor-DNN}, respectively. 

\begin{figure}[!t]
    \centering
    \includegraphics[width=0.85\columnwidth]{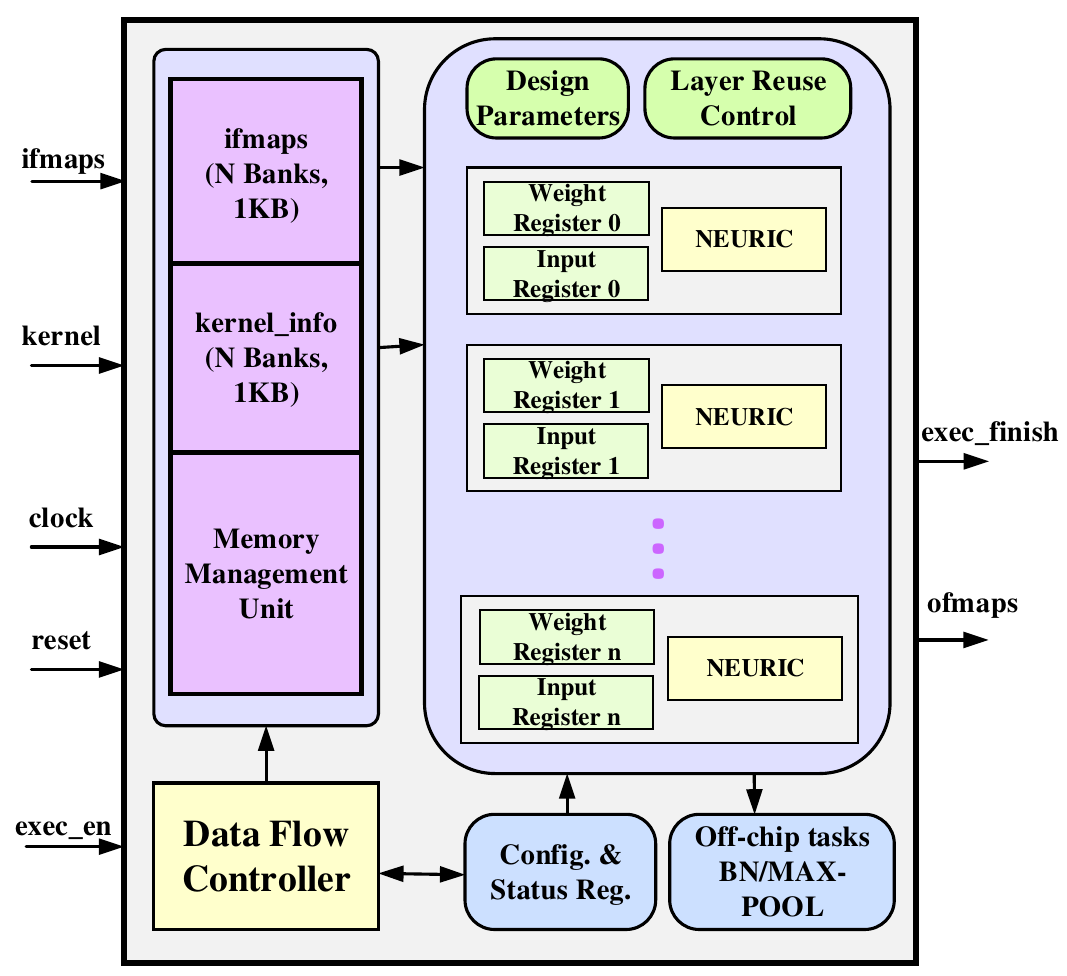}
    \vspace{-3mm}
    \caption{Performance-Enhanced Dynamically Configurable Layer Multiplexed Vector Engine for AI Acceleration.}
    \vspace{-3mm}
    \label{fig:smart-af-arch}
\end{figure}

\begin{figure*}
    \centering
    \includegraphics[width=0.88\linewidth]{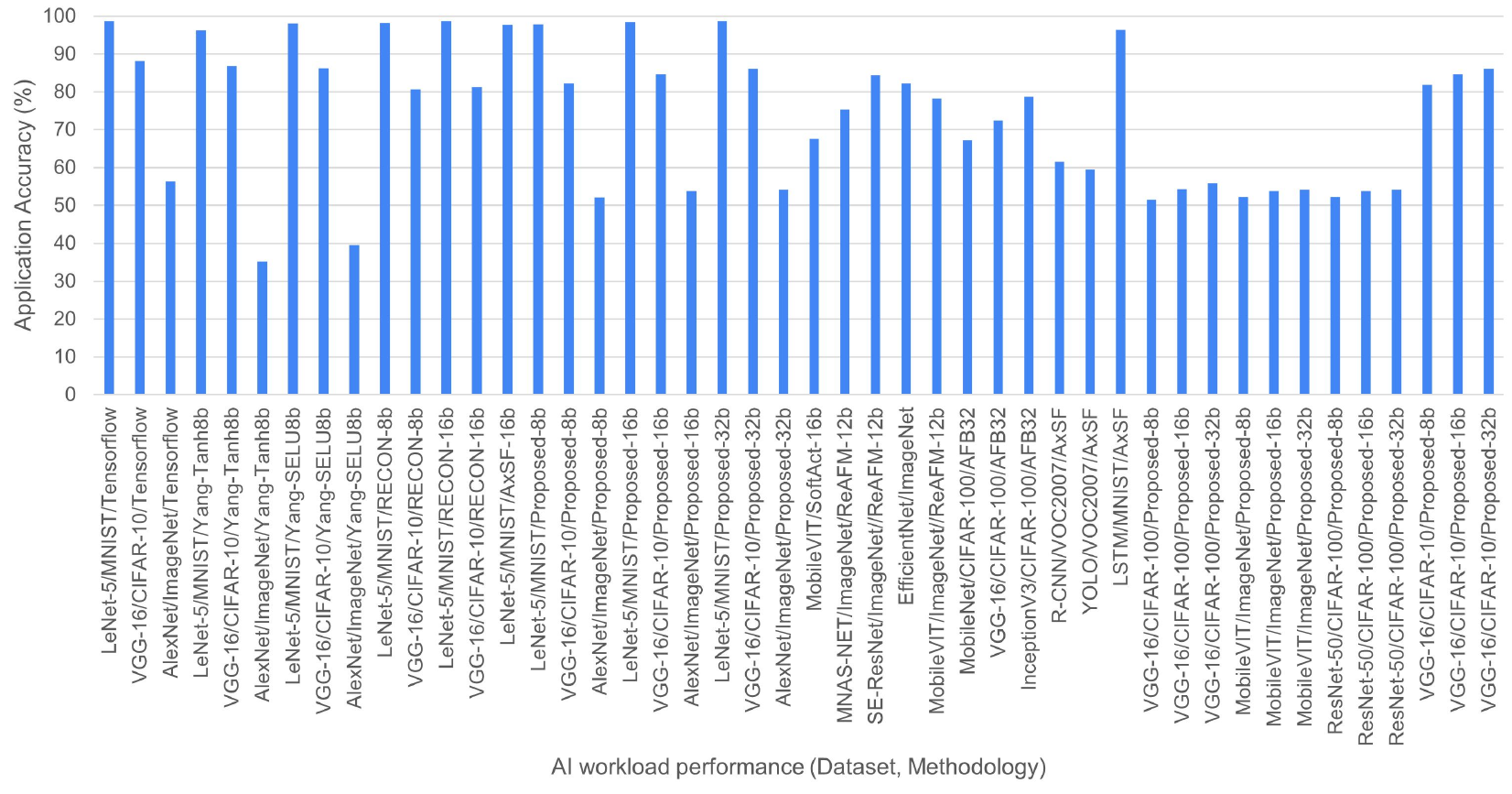}
    \vspace{-3mm}
    \caption{Comparison of AI application accuracy evaluation with state-of-the-art works~\cite{Designspaceexploration-AF, RECON, SoftMax-taylor-DNN, SoftAct-Trans, ReAFM-NN, CORDICAF-LSTM}.}
    \vspace{-3mm}
    \label{perf-eval}
\end{figure*}

The error metrics were calculated with functional emulation for the proposed hardware architecture with Python v3.0 and Numpy true value for corresponding floating-point precision and found to be around 5-10\%. 
Error metrics assist designers in understanding the degree of approximation induced in neural networks and exploring performance-quality trade-offs. 
The error metrics were compared with SoTA works for 8/16/32-bit precision design as shown in \figref{fig:smart-af-error}. We observed a significant reduction in Mean error and Mean Absolute Error (MAE) as shown in \figref{fig:smart-af-error}. 
This reflects an improvement in application accuracy, as detailed in \figref{perf-eval}. 
The definition for these metrics are:

\begin{equation}
    \begin{aligned}
\text{ME} = \frac{1}{N} \sum_{i=1}^{N} (Acc\_Y_i - Pred\_\hat{Y_i}) \\
\text{MAE} = \frac{1}{N} \sum_{i=1}^{N} |Acc\_Y_i - Pred\_\hat{Y_i}| 
    \end{aligned}
    \label{error-eq}
\end{equation}

\begin{table}[!t]
\caption{Comparison of FPGA resources: Ac2Ap2-PE~\cite{Acc-App-PE}, QuantPE~\cite{Quant-PE}, Xilinx-IP(Xil-PE)~\cite{GR-ACM_TRETS23}, FP-MPE, FP-RPE~\cite{Reconfigurable-PE-TCASII24}, with NEURIC.}
\label{table:comp-PE}
\renewcommand{\arraystretch}{1.4}
\resizebox{\linewidth}{!}{%
\begin{tabular}{c|c|c|c|c|c|c}
\hline
\textbf{Parameters} & \textbf{Ac2Ap2-PE~\cite{Acc-App-PE}} & \textbf{QuantPE~\cite{Quant-PE}} & \textbf{Xil-PE~\cite{GR-ACM_TRETS23}} & \textbf{FP-MPE~\cite{Reconfigurable-PE-TCASII24}} & \textbf{FP-RPE~\cite{Reconfigurable-PE-TCASII24}} & \textbf{NEURIC} \\ \hline
\textbf{LUTs} & 299 & 463 & 511 & 8167 & 8234 & 537 \\ \hline
\textbf{FFs} & 367 & 298 & 357 & 1284 & 1918 & 468 \\ \hline
\textbf{Delay (ns)} & 6.69 & 5.96 & 8.244 & 5.72 & 5.27 & 2.56 \\ \hline
\textbf{Power (mW)} & 142.4 & 45.3 & 59.7 & 384 & 298.7 & 30.38 \\ \hline
\textbf{\begin{tabular}[c]{@{}c@{}} Arith. Intensity \\ (pJ/Op)\end{tabular}} & 954 & 270 & 491 & 2196 & 1583 & 133 \\ \hline
\end{tabular}}
\vspace{-2mm}
\end{table}

The proposed design was synthesized using a CMOS 28\,nm HPC+ (0.9V) and 45\,nm nodes with Synopsys Design Compiler, and the post-synthesis ASIC parameters are compared with the SoTA designs as reported in \tblref{tab:ASIC-comp}. 
The proposed design provides an area reduction of up to 16.2$\times$ and 4$\times$ while optimizing the critical delay to 7.8$\times$ and 1.3$\times$ compared to the SoTA works~\cite{TCASI23-Softmax} and~\cite{RECON}, respectively. 
This comes with a 2.1$\times$ and 14.3$\times$ reduction in power, as compared to~\cite{CORDICAF-LSTM} and~\cite{RECON}, respectively, thus enhancing the run time of battery-powered AI devices. 
Furthermore, the design reduces up to 11.5$\times$ area, 1.75$\times$ critical delay, and 17.3$\times$ power consumption compared to RECON~\cite{RECON}. Additionally, it shrinks 2.5$\times$ power and 1.8$\times$ area reduction over~\cite{CORDICAF-LSTM} and~\cite{SoftMax-taylor-DNN}, respectively, at the CMOS 28\,nm node. 
The performance analysis showcases the proposed solution is a balanced approach with significant enhancement compared to prior works~\cite{RECON, TCASI23-Softmax}, as shown in \figref{fig:perf-comp}. 
The core delivers enhanced compute density and is well-suited for resource-constrained edge devices, at the CMOS 28\,nm node.

\begin{table}[!t]
    \caption{Comparative analysis for FPGA Resources (VC707) with Layer-Multiplexed~\cite{GR-ACM_TRETS23} Vector Engine (Benchmark Arch. VGG-16/CIFAR-100) with SoTA PE designs}
    \label{HW-arch-PE16}
    \renewcommand{\arraystretch}{1.4}
    \resizebox{\columnwidth}{!}{%
    \begin{tabular}{l|r|r|r|r|c|r|r}
    \hline
\textbf{Design} & \textbf{Precision} & \textbf{k-LUTs} & \textbf{k-Regs/FFs} & \textbf{DSPs} & \textbf{Op. Freq. (MHz)} & \textbf{GOPS/W} & \textbf{Power (W)} \\ \hline

\textbf{Xil-PE}~\cite{GR-ACM_TRETS23} & 16 & 90.6 & 51.7 & 36 & 100 & 3.1 & 5.8 \\\hline
\textbf{QuantPE}~\cite{Quant-PE} & 8 & 69.2 & 75.4 & 58 & 200 & 2.7 & 6.7 \\\hline
\textbf{Ac2Ap2-PE}~\cite{Acc-App-PE} & 16 & 66.3 & 75.4 & 108 & 150 & 3.6 & 3.6 \\\hline
% \textbf{RPE}~\cite{Reconfigurable-PE-TCASII24} & 16 & 310 & 126 & 492 & 216 & 3 & 21.74 \\\hline
\textbf{CORDIC-Neuron}~\cite{GR-ACM_TRETS23} & 8 & 144 & 155 & 23 & 50 & 5 & 2.24 \\\hline
\textbf{NEURIC} & 8/16 & 53 & 31.2 & 112 & 354 & 8.4 & 1.54 \\ \hline

\end{tabular}%
}
\end{table}

\subsection{Performance-Enhanced Vector Engine }
To address the enhanced throughput requirements in AI hardware, Vector Engine are the important focus for optimization. We analyzed NEURIC-based programmable dynamically configurable Vector Engine with 64 NEURIC units and compared the performance with similar SoTA designs, as shown in \figref{fig:smart-af-arch}.
The proposed accelerator achieves an inference time of 148 ms providing a throughput of 52.3 GOPS compared to 184 ms and 42.1 GOPS of posit-TREA architecture~\cite{LPRE}. 
While, fixed-point-Layer reused architecture~\cite{GR-ACM_TRETS23} achieves 772 ms at 4.95 GOPS throughput on \textit{Xilinx Virtex-7} FPGA Board and 226 mili-seconds with 34.38 GOPS on \textit{Jetson Nano} at INT/FxP8 precision with an accuracy of 95\%. 
Therefore, the proposed approach provides an efficient trade-off between available hardware resources and throughput latency with run-time programmability feature. 
The performance enhancement in Transformers with the proposed FxP8 NEURIC achieves an improvement in energy efficiency up to 2.5\X over TRETA~\cite{Transformers-Training-TVLSI'23} and 17.54\X  over SIGMA~\cite{Transformers-Training-TVLSI'23}. 
The power consumption also drops significantly, compared to TRETA (4.45 W) and SIGMA (22.33 W) for MobileViT/ImageNet. 
Our design achieves a speed boost up to 1.8\X with 2.33\% improved accuracy compared to TRETA~\cite{Transformers-Training-TVLSI'23}. 
NEURIC provides this additional performance enhancement with almost the same resources for AF blocks and Matrix Multiplication. 
We leave the detailed exploration of NEURIC for diverse AI workloads as future work.

\section{Summary}
\label{sec_conclusions}
 %    \ExecuteMetaData[Templates/templateText]{conclusions}
Depending on AI workloads, DA-VINCI, a dynamically configurable and precision-scalable AF core, can be configured to implement the following AFs: Swish, SoftMax, SELU, GELU, Sigmoid, Tanh, and ReLU. 
%CORDIC-based configurable AF can not be written off. 
The proposed NEURIC-based Vector Engine approach even shows enhanced resource efficiency and compute density with 98.5\% QoR. 
Thus, our resource-efficient design solution is viable for low-power, high-performance AI accelerators targeting DNNs, RNNs/LSTMs, and Transformer applications. 

%%%%%%%%%%%%%%%%%%%%%%%%%%%%%%%%%%%%%%%%%%%%%%%%%%%%%%%%%%%%%
%    This file configures your bibliography
%        Just a bit cleaner than having a long list of bib 
%          files in your main document
%%%%%%%%%%%%%%%%%%%%%%%%%%%%%%%%%%%%%%%%%%%%%%%%%%%%%%%%%%%%%

% Start by defining the bibliography style
%    Some commonly used styles:
%       IEEEtran    - for most IEEE Transactions and Conference Proceedings

\ifmicro
    \bibliographystyle{IEEEtranS}
\else
    \bibliographystyle{IEEEtran}
\fi

% Now provide a list of bibliography files to include within the \bibliography command
%   These files are stored in the "bibliography" folder
%   There is an "abbreviations.bib" file which includes long and short names for common journals and conferences
%   There should be a file for papers published by each of the EnICS Staff members
%   There is a "general_bibliography" file for things we tend to cite a lot (e.g., ITRS)
%   Put your paper-specific citations in the "this_bibliography.bib" file
\bibliography{bibliography/abbreviations,
              bibliography/general_biblography,
              bibliography/teman_bibliography,
              bibliography/this_bibliography}

%\bibliography{bibliography/this_bibliography}
\end{document}